\documentclass[12pt, Letter]{article}

\begin{document}
\date{}
\title{Comments on ``Improved Efficient Remote User Authentication Schemes''}
\author{Manik Lal Das\\
\emph{Dhirubhai Ambani Institute of Information and Communication
Technology}\\
\emph{Gandhinagar - 382007, Gujarat, India.}\\
\emph{Email: maniklal\_das@daiict.ac.in}}

\maketitle

\begin{abstract}
Recently, Tian et al. presented an article, in which they
discussed some security weaknesses of Yoon et al.'s scheme and
subsequently proposed two ``improved'' schemes. In this paper, we
show that the Tian et al.'s schemes are insecure and vulnerable
than the Yoon et al.'s scheme.\vspace{2 mm}
\end{abstract}

\section{Introduction}
Remote system authentication is a process by which a remote system
gains confidence about the identity (or login request) of the
communicating partner. Since the introduction of Lamport's scheme
\cite{lam81}, several new proposals and improvements on remote
systems authentication \cite{jab96}, \cite{hwa00}, \cite{lee02},
\cite{awa04}, \cite{das04} have been proposed. Recently, Tian et
al. \cite{tia07} presented an article by observing some flaws of
the Yoon et al.'s scheme \cite{yoo04}, and subsequently suggested
two improved schemes. The basis of the Tian et al.'s observation
on Yoon et al.'s scheme was on this assumption: \textit{If an
attacker steals a user's smart card and extracts the values stored
in it through some means \cite{koc99}, \cite{mes02} without being
noticed, then the attacker can either masquerade as the user to
forge a valid login request, or masquerade as the server to forge
a valid reply message}.\\
In this paper, we show that the Tian et al.'s schemes are insecure
with the above mentioned arguments what they had considered, in
fact, more vulnerable than \cite{yoo04}. The remainder of the
paper is organized as follows. In the next section, we review the
Tian et al.'s schemes. In section 3, we show the security
weaknesses of the schemes. We conclude the paper with the section
4.

\section{The Tian et al.'s Schemes}
The schemes consists of four phases: Registration, Login,
Authentication and Password change. The registration and password
change phases are same for both the schemes.\vspace{2 mm}\\
\textbf{Registration phase}: A new user can register to the remote
server by the following steps.
\newcounter{3}
\begin{list}{R\arabic{3}.}
{\usecounter{3}} \item A user $U_i$ submits his identity $ID_i$
and password $PW_i$ to the server ($S$) through a secure channel.
\item Then $S$ chooses four distinct cryptographic one-way hash
functions $h(\cdot)$, $h_1(\cdot)$, $h_2(\cdot)$, and
$h_3(\cdot)$. \item $S$ computes $R_i = h(ID_i, x_s)$, $H_i =
h(R_i)$ and $X_i = R_i \oplus h(ID_i, PW_i)$, where $\oplus$
denotes the bitwise exclusive-OR operation. \item Then $S$
personalizes a smart card with $< ID_i, H_i, X_i, h(\cdot),
h_1(\cdot), h_2(\cdot), h_3(\cdot) >$ and sends it to $U_i$ in a
secure manner.
\end{list}
\textbf{Password change phase}: This phase is invoked when a user
$U_i$ wants to change his password from $PW_i$ to $PW_i^{\prime}$.
The user attaches his smart card to the card reader and enters
$PW_i$, then the smart card performs the following operations:
\newcounter{c}
\begin{list}{P\arabic{c}.}
{\usecounter{c}} \item Compute $R_i^{\prime} = X_i \oplus h(ID_i,
PW_i)$ and $H_i^{\prime} = h(R_i^{\prime})$. \item Compare
$H_i^{\prime}$ with $H_i$. If they are equal, then the user enters
a new password $PW_i^{\prime}$, otherwise it rejects the password
change request. \item Compute $X_i^{\prime} = R_i \oplus h(ID_i,
PW_i^{\prime})$. Then, store $X_i^{\prime}$ in smart card in place
of $X_i$.
\end{list}

\subsection{The First Scheme}
This scheme uses the timestamp mechanism to avoid the replay
attack (assuming the user and server time synchronization is proper).\vspace{2 mm}\\
\textbf{Login phase}: $U_i$ attaches his smart card to the card
reader and enters password $PW_i^*$. Then the smart card performs
the following operations:
\newcounter{2}
\begin{list}{LF\arabic{2}.}
{\usecounter{2}} \item Compute $R_i^{\prime} = X_i \oplus h(ID_i,
PW_i^*)$ and $H_i^{\prime} = h(R_i^{\prime})$. \item Compare
$H_i^{\prime}$ with $H_i$. If they are equal, then the smart card
proceeds to the next step, otherwise it terminates the operation.
\item Compute $C_1 = h_1(S, ID_i, R_i, T)$, where $T$ is the
timestamp. \item $U_i$ sends the login request $< ID_i, T, C_1>$
to $S$ over a public channel.
\end{list}
\textbf{Authentication phase}: Upon receiving the login request $<
ID_i, T, C_1 >$, the server $S$ and the user $U_i$ perform the
following steps for mutual authentication:
\newcounter{1}
\begin{list}{AF\arabic{1}.}
{\usecounter{1}} \item $S$ checks the validity of $ID_i$ and $T$.
If both are correct then proceeds to the next step, otherwise
rejects the login request. \item $S$ computes $R_i = h(ID_i, x_s)$
and checks whether $C_1 = h_1(S, ID_i, R_i, T)$. If this check
holds, $S$ assures that $U_i$ is authentic and proceeds to the
next step, otherwise it rejects the request. \item $S$ computes
$C_2 = h_2(ID_i, S, R_i, T^{\prime})$, where $T^{\prime}$ is a
timestamp. Then, $S$ sends $< T^{\prime}, C_2 >$ back to $U_i$
through the public channel. \item Upon receiving $S$'s response
message $< T^{\prime}, C_2 >$, $U_i$'s smart card first checks the
validity of $T^{\prime}$ and then whether $C_2 = h_2(ID_i, S, R_i,
T^{\prime})$. If these checks hold, $U_i$ assures the authenticity
of $S$ and the mutual authentication is done, otherwise it rejects
the connection. \item Once the mutual authentication is completed,
$U_i$ and $S$ use $h_3(ID_i, S, R_i, T, T^{\prime})$ as the
session key.
\end{list}

\subsection{The Second Scheme}
This scheme uses a nonce based challenge-response mechanism, so it
avoids the time synchronization problem.\vspace{1 mm}\\
\textbf{Login phase}: $U_i$ attaches his smart card to the card
reader and enters password $PW_i$. Then the smart card performs
the following operations:
\newcounter{b}
\begin{list}{LS\arabic{b}.}
{\usecounter{b}} \item Compute $R_i^{\prime} = X_i \oplus h(ID_i,
PW_i)$ and $H_i^{\prime} = h(R_i^{\prime})$. \item Compare
$H_i^{\prime}$ with $H_i$. If they are equal, proceeds to the next
step, otherwise it terminates the operation. \item Send the login
request $< ID_i, N_i >$ to $S$ over a public channel, where $N_i$
is a nonce selected by $U_i$.
\end{list}
\textbf{Authentication phase}: Upon receiving the login request $<
ID_i, N_i >$, the server $S$ and the user $U_i$ perform the
following steps for mutual authentication:
\newcounter{a}
\begin{list}{AS\arabic{a}.}
{\usecounter{a}} \item $S$ checks the validity of $ID_i$. \item
$S$ chooses a nonce $N_s$, computes $R_i = h(ID_i, x_s)$, $C_1 =
h_1(S, ID_i, R_i, N_i, N_s)$ and sends $< C_1, N_s >$ to $U_i$
over a public channel. \item Upon receiving $< C_1, N_s >$, $U_i$
checks whether $C_1 = h_1(S, ID_i, R_i, N_i, N_s)$. If this check
holds correct, $U_i$ assures the authenticity of $S$, otherwise
terminates the operation. \item $U_i$ computes $C_2 = h_2(ID_i, S,
R_i, N_s, N_i)$ and sends it to $S$. \item Upon receiving $C_2$,
$S$ checks whether $C_2 = h_2(ID_i, S, R_i, N_s, N_i)$. $U_i$
authentic if the check passes and the mutual authentication is
done, otherwise $S$ terminates the operation. \item After the
mutual authentication, the user and the server use $h_3(ID_i, S,
R_i, N_i, N_s)$ as the session key.
\end{list}

\section{Security Weaknesses}
The basis of the following attacks is based on this risk of smart
card stored information:\\
\textit{A legitimate user could extract the values stored in smart
card by some means \cite{koc99}, \cite{mes02}, then he/she could
act as the role of server to register any number of users. We note
that the Tian et al.'s scheme also assumed a similar
risk}.\vspace{1
mm}\\
1. \textbf{Attacks by a legitimate user}\\
In the registration phase, $X_i = R_i \oplus h(ID_i, PW_i)$ is
stored in $U_i$'s smart card. Once $U_i$ extracts $X_i$ from his
smart card by some means \cite{koc99}, \cite{mes02}, then he/she
can easily get $R_i$ by computing $R_i = X_i \oplus h(ID_i,
PW_i)$. After that, no remote server is required to register a new
user. Now, $U_i$ who has $R_i$, could register any number of users
by distributing $R_i$ and $ID_i$. In fact, smart card and password
are not required at all to login $S$ those who got $R_i$ and
$ID_i$ from $U_i$. Because, a valid login message is $< ID_i, T,
C_1 >$, where $T$ is a timestamp (for the first scheme) and $C_1 =
h_1(S, ID_i, R_i, T)$. For the second scheme, the
challenge-response comprises with the secret $R_i$ only, other
parameters are public. Therefore, the server secret is
virtually compromised by a legitimate user's smart card.\vspace{1 mm}\\
2. \textbf{Attacks by an adversary}\\
Suppose an attacker steals $U_i$'s smart card and intercepts $C_1
= (S, ID_i, R_i, T)$ from a valid login request. Now the attacker
extracts the information stored in the smart card and launches an
offline guessing attacks of $PW_i$ in order to obtain the value of
$R_i$. The attacker guesses a password and obtains an $R_i^*$, and
then checks whether $C_1 = h_1(S ID_i, R_i^*, T)$. Once the guess
succeeds, then the attacker
has a valid $R_i$ and can create any number of valid login request.\vspace{2 mm}\\
3. \textbf{No two-factor authentication}\\
Two-factor authentication is a technique that requires two
independent factors (e.g. password, smart card) to establish
identity and privileges. Common implementations of two-factor
authentication use `something you know: password' as one of the
two factors, and use either `something you have: smart card' or
`something you are: biometric' as the other factor. A common
example of two-factor authentication is a bank card (credit card,
debit card); the card itself is the physical item, and the
personal identification number (PIN) is the data that goes with
it.\\
In Tian et al.'s scheme, we observe that once a party has
information of $ID_i$ and $R_i$, then he does not require password
and a valid smart card at all. Without password and smart card,
one can easily pass the mutual authentication and establish the
session key. Therefore, the schemes lack two-factor
authentication.

\section{Conclusion}
The threat of smart card security \cite{koc99}, \cite{mes02},
\cite{joy05} is a crucial concern, where some secret information
is stored in the memory of smart cards. However, to the best of my
knowledge, one can still use smart card to store some secret data
by considering the applications requirement and scope/value of the
secret information stored in the smart card. It is also important
to judge the financial cost and time to extract the secret data
from the smart card. If the cost as well as time is tolerable or
higher than the cost of the secret inside the smart card, then one
can take that risk while using smart card to store some secret
data. If extracting a secret from the card leads to collapse the
whole system (e.g. Tian et al.'s schemes) then definitely some
additional counter measure should be taken while designing the
scheme. Of course, smart card vendors are quite aware of these
threats and they are also taking counter measure continuously to
safe guard the cards security.\\
We have shown that the Tian et al.'s scheme is insecure by several
weaknesses. Just by extracting a secret data from a smart card can
collapse the whole system's security.

\end{document}